\documentclass[a4paper,twocolumn,
english,aps,pre,floatfix,showpacs,superscriptaddress]{revtex4}
\usepackage[T1]{fontenc}
\usepackage[latin1]{inputenc}
\usepackage{amsmath}
\usepackage{babel}
\usepackage{graphics}
\usepackage{amssymb}
\usepackage{color}  


\begin{document}

\title {Finite-size scaling behavior in trapped systems}

\author {S. L. A. \surname{de Queiroz}}
\email{sldq@if.ufrj.br}
\author {R. R. \surname{dos Santos}}
\email{rrds@if.ufrj.br}
\affiliation{Instituto de F\'\i sica, Universidade Federal do Rio de Janeiro, Caixa
Postal 68528, 21941-972 Rio de Janeiro RJ, Brazil}


\author {R. B. \surname{Stinchcombe}}

\email{r.stinchcombe1@physics.ox.ac.uk}

\affiliation{Rudolf Peierls Centre for Theoretical Physics, University of Oxford, 1 Keble
Road, Oxford OX1 3NP, United Kingdom}


\begin{abstract}
Numerical transfer-matrix methods are applied to two-dimensional 
Ising spin systems, in presence of a confining magnetic field
which varies with distance $|{\vec x}|$ to a "trap center", proportionally to
$(|{\vec x}|/\ell)^p$, $p>0$. 
On a strip geometry, the competition between the
"trap size"  $\ell$ and the strip width, $L$, is analysed in the context
of a generalized finite-size scaling {\em ansatz}. In the low-field 
regime $\ell \gg L$, we use conformal-invariance concepts in conjunction with 
a linear-response approach to derive the appropriate ($p$-dependent) 
limit of the theory, 
which agrees very well with numerical results for magnetization profiles. 
For high fields $\ell \lesssim L$,
correlation-length scaling data broadly confirms an existing picture of 
$p$-dependent characteristic exponents. Standard spin-$1/2$ and spin-$1$ Ising 
systems are considered, as well as the Blume-Capel model. 
\end{abstract}

\pacs{05.70.Jk,64.60.F-,67.85.-d }

\maketitle 

\section{Introduction} \label{intro}

The continuing progress in the field of trapped atomic quantum gases~\cite{cw02,k02}
has led to the development of a variety of experimental techniques (e.g.,
{\it in situ} imaging~\cite{Schneider08}) to investigate Mott insulating phases in
trapped bosonic or fermionic atoms~\cite{Bloch08}.
Inspired by the physics of strongly-correlated systems, a further step in the
agenda will be the investigation of spin-ordered phases, given the recent
improvements on cooling methods and ultra--low-temperature measurements.
An important ingredient of the latter advance is the nature and shape of the
atomic density profile within the
trap, leading to a free gas at the wings of the distribution~\cite{Kohl06,Ketterle09}.
As the accuracy of these measurements improves, a detailed description of
the possible crossovers involved will be needed in order to analyze the
experimental data.
Indeed, for an atomic sample of finite size, $L$, the presence of a 
trap brings about another length scale, $\ell$ (to be defined more precisely below),
characterizing the trap shape and size.
Therefore, the competition between these (geometric) lengths and the one describing
the decay of correlations, $\xi$, which is assumed to diverge at the
ordering
phase transition in the thermodynamic limit, must be incorporated in a
new scaling theory, which generalizes the standard finite-size scaling
(FSS)~\cite{barber}.
We should mention that a {\em trap-size} scaling regime, in which
both $\xi$ and $\ell$ are large,
has been considered in Ref.~\cite{cv09}, with the 
relevant parameters adjusted in such a way
that $\ell < L$, thus
eliminating most finite sample-size effects.

Our purpose here is to develop such generalized scaling theory, and we also adopt the
strategy of leaving aside all quantum effects~\cite{cv09b} and concentrating instead on
classical spins localized on sites of a lattice, subject to a trapping field.
More specifically, we consider Ising-like spins under a
spatially varying magnetic field, whose intensity increases with distance $|{\vec x}|$ to
the origin (trap center)~\cite{pbs08,cv09}. This way, for suitably large $|{\vec x}|$ the
spins are pinned parallel to the local field. With the well-known mapping of the
(pseudo)-spin variable onto a lattice-gas picture, one thus gets at least the main
qualitative features of trapping phenomena.

We investigate the scaling behavior of two-dimensional Ising spins close to
the critical temperature $T_c$ at which, in zero field and in the thermodynamic limit,
the correlation length $\xi$ becomes much larger than the lattice parameter, $a$ (to be
taken as unity in what follows).
Initially, only spin$-1/2$ systems are considered. 
Upon addition of a trapping magnetic field, as outlined above, $\xi$ is prevented from diverging, thus smearing out the second-order phase
transition.
Nonetheless, we can still look for signatures of the smeared phase transition in 
both the trap scaling regime $\ell \lesssim L$, considered in Ref.~\onlinecite{cv09}, 
as well as in a "shallow trap" regime, in which $\ell\gg L$.

In both regimes we use numerical transfer matrix (TM) methods and FSS ideas to 
disentangle the respective trapping and finite-size effects in Ising-like systems 
which are necessarily finite in extent. The quantities of interest, such as free 
energies, equilibrium site magnetizations, bond energies, correlation functions 
and correlation lengths, can be extracted from the leading eigenvalues of the TM, 
and from their respective eigenvectors~\cite{night}. As we will see, in the 
shallow trap regime, conformal invariance allows us to develop a perturbative 
theory, whose predictions are in excellent agreement with the numerical data. 
We investigate the quantitative dependence of scaling behavior on trap shape, 
and test for universality against spin magnitude, the latter by 
considering both a standard spin-1 Ising system and the Blume-Capel model.

The layout of the paper is as follows.  In Sec.~\ref{sec:basics} we parametrize 
the trapping field in more detail, and present the scaling {\it ansatz} for the 
free energy, making contact with previous work \cite{cv09}, when applicable; 
highlights on some technical aspects of the methods used in this paper are also 
given. Section~\ref{sec:lowfield} deals with the low-field (or shallow trap) 
regime, in which scaling predictions are drawn perturbatively from conformal 
invariance, and checked against numerical TM data. TM methods are used to 
probe the high-field (or steep trap) regime in Secs.~\ref{sec:hf} and 
\ref{sec:bc}, the latter concerning the behavior of $S=1$ spins. Finally, in 
Sec.~\ref{sec:conc}, concluding remarks are made.

\section{Basic aspects} \label{sec:basics}

We consider a trapping field with a single power-law dependence on distance to the trap
center:
\begin{equation}
h({\vec x}) = h_0\,|{\vec x}|^{\,p} \equiv \left(|{\vec x}|/\ell\,\right)^p\ .
\label{eq:field_def}
\end{equation}
Thus $\ell =h_0^{-1/p}$ is an effective "trap size".
In cold-atom experiments, harmonic potentials ($p=2$) are usual.

We use TM methods on long strips of width $L$ sites of a square lattice, with periodic
boundary conditions (PBC) across. So $\ell \gg L$ corresponds, for all $\vec x$ inside
the strip, to a low-field regime where at $T$ near $T_c$ and large $L$ the system is
still in the scaling regime (Section~\ref{sec:lowfield}); and the same is true
at $|{\vec x}|$ near zero, even for $\ell ={\cal O}(L)$ ("high fields",
Section~\ref{sec:hf}).
In the strip geometry, the trapping field is translationally
invariant along the "infinite" direction; given Eq.~(\ref{eq:field_def}), consistency
with PBC demands that the field be symmetric relative to the line halfway along the strip
width, thus in this case the distance $|{\vec x}|$ of Eq.~(\ref{eq:field_def}) stands for
position relative to such an axis. It is expected~\cite{pbs08,watb04}, and has been
verified numerically~\cite{cv09}, that the scaling properties of trapping in this
one-dimensional well will be the same as in a full two-dimensional one, for which the
equipotentials are concentric circumferences.

With $t \equiv (T-T_c)/T_c$, and a uniform magnetic field $H$, a FSS expression for the
singular part of the free energy  can be adapted to the trapping context, as follows:
\begin{equation}
F(t,H,\ell,L) =b^{-d}\,F(t\,b^{\,y_t},H\,b^{\,y_H},\ell^{-1}
b^{\,y_\ell},L^{-1} b^{\,y_L})\ ,
\label{eq:fss_fe}
\end{equation}
where $b$ is an
arbitrary rescaling parameter; $y_t=1/\nu$, $y_H=(d+2-\eta)/2$ are the usual scaling
exponents; $y_L=1$, and the "trapping" exponent $y_\ell$ is given, in this case where the
trapping field couples to the magnetization, by~\cite{cv09}:
\begin{equation}
\frac{1}{y_\ell} \equiv \theta = \frac{2p}{d+2-\eta+2p}\ .
\label{eq:theta}
\end{equation}
Making $b=L$, one gets:
\begin{equation}
F(t,H,L,\ell)=L^{-d}\,{\cal F}\left(L\,t^\nu,H\,L^{y_H},L\,\ell^{-\theta}\right)
\label{eq:fss_fe2}
\end{equation}

This prescribes the scaling behavior, in the trap, of all thermodynamic properties.
Beyond scaling, their specific forms are not provided by Eq.~(\ref{eq:fss_fe2}), and they
will be modified by the trap.

\section{Low-field regime}
\label{sec:lowfield}
The trapped system we consider is a
spin-$1/2$ Ising one, (spin variables $\sigma=\pm 1$ on sites of a square lattice) with
nearest-neighbor interactions of unit  value, so the exact critical point is at
$T_c=2/\ln(1+\sqrt{2})$, $H=\ell^{-1}=L^{-1}=0$. Setting $t=H=0$, we first focus on the
low-field regime corresponding to $\ell \gg L$. In renormalization-group terminology, one
has two relevant nonzero fields, namely $L^{-1}$ and $\ell^{-1}$. As the effects of the
first of these are well-understood via conformal invariance~\cite{cardy}, they can be
readily incorporated into our description. In the regime under consideration, the
resulting picture is a suitable starting point for a perturbative treatment of the second
field.

Within a linear-response context, and with aid of the fluctuation-dissipation theorem,
one can write for the equilibrium magnetization at site $i$:
\begin{equation}
\langle\sigma_i\rangle = \sum_j \langle \sigma_i\,\sigma_j \rangle_c\,h_j ,
\label{eq:lin-resp}
\end{equation}
where $\langle \cdots\,\rangle$ denotes thermodynamic average, $h_j$
represents the trapping field at site $j$, and the subscript $c$ stands for connected
spin-spin correlation functions.

In keeping to the first-order perturbation picture adopted here, one substitutes the
zero-field expressions, $C_{ij}\equiv\langle \sigma_i\,\sigma_j \rangle^0_c$ on the
right-hand side of Eq.~(\ref{eq:lin-resp}). These, in turn, are given via conformal
invariance~\cite{cardy}, on a strip of width $L$ with PBC, by:
\begin{equation}
C_{xy} \sim  \frac{(2\pi/L)^\eta}{\left[ 2\cosh (2\pi x/L)-2 \cos
(2\pi y/L)\right]^{\,\eta/2}}\  ,\  \eta =1/4\ .
\label{eq:conf-inv}
\end{equation}
Here, $x$,  $y$ are the relative
spin-spin coordinates, respectively along the strip, and across it.

Though strictly speaking, Eq.~(\ref{eq:conf-inv}) is an asymptotic form, discrepancies
are already very small at short distances, amounting to less than $0.5\%$ for separations
of the order of three lattice spacings~\cite{dqrbs03,dq06}.

Setting the origin at a point on the trapping axis, thus $h_j \equiv
h(x,y)=h_0\,|y|^{p}$, using Eq.~(\ref{eq:conf-inv}) in Eq.~(\ref{eq:lin-resp}), and
transforming  the latter into an integral, one gets:
\begin{equation}
\langle\sigma_0\rangle \equiv \langle \sigma (y=0)\rangle =
h_0\,\left(\frac{L}{2\pi}\right)^{2+p-\eta}\,{\cal A}(p)\ ,
\label{eq:int-a}
\end{equation}
\begin{equation}
{\cal A}(p)\ = A_1(p)+A_2(p)\ ,
\label{eq:int-b}
\end{equation}
where $A_1(p) $ represents the long-distance contribution ($x/L > 1$, say)
for which it is justified to ignore the $y$-- dependence of Eq.~(\ref{eq:conf-inv}), and
$A_2(p)$ takes fully into account the short-range, angle-dependent, part of the integral.
For $x$, $y \ll L$ in the latter, the denominator in Eq.~(\ref{eq:conf-inv}) is
proportional to $R^2=x^2+y^2$, and one can use polar coordinates for the integration
close to the origin. Further contributions of order $1/L$ to ${\cal A}(p)$ have been
neglected.

One finds:
\begin{eqnarray}
A_1(p) =\frac{4}{\eta\pi}\,e^{\eta\pi/2}\,\frac{\pi^{\,p+1}}{p+1}\ ;
\label{eq:a1} \\
A_2(p) =\frac{1}{\pi}\,\frac{(\alpha\pi)^{\,p-\eta+2}}{p-\eta+2}\,C(p)\ ,
\label{eq:a2}
\end{eqnarray}
where $\alpha$ is a constant of order unity (reflecting the choice of
region where polar coordinates are used), and 
$C(p) =\int_0^{\pi/2}
(\sin\theta)^p\,d\theta=[\,(p-1)!!\,/\,p!!\,]\,a$, where $a=1$ for odd $p$,
and $\pi/2$ for even $p$.


The leading scaling behavior of the magnetization can be
extracted from Eq.~(\ref{eq:fss_fe2}), by first
replacing $\ell=h_0^{-1/p}$ there, thus showing that  
$h_0$ and $L$ always occur in the combination
$h_0\,L^{\,p/\theta}$ in the free energy. 
However, as can be seen from Eq.~(\ref{eq:field_def}),
the field amplitude $h_0$ does not 
have the dimension of a magnetic field. Accordingly, given the
non-uniform character of $h({\vec r})$, the appropriate 
field variable  (with the correct dimension) to
use when differentiating the free energy to obtain the magnetization 
is a spatially averaged, coarse grained, one; for the strip 
geometry considered here it reads, 
\begin{equation} 
{\overline h} \sim {1 \over L} 
\int_{-L/2}^{L/2} h_0\,|\,y |^{\,p}\,dy \sim h_0\,L^{\,p}\ . 
\label{eq:h-ave}
\end{equation} 
Finally, within a linear response approach, a   
linear dependence of the magnetization on field corresponds to a     
quadratic dependence of the free energy. We then have $F \sim L^{-d}\
\left( h_0 L^{\,p/\theta}\right)^2$, which yields 
\begin{equation}
\langle \sigma \rangle \sim \frac{\partial F}{\partial {\overline h}}
=\frac{1}{L^{\,p}}\frac{\partial F}{\partial h_0} \sim h_0 L^{\,2+p-\eta}\ ,
\label{eq:M} 
\end{equation} 
in agreement with Eq.~(\ref{eq:int-a}).
Note further that the above derivation is valid also away from the central 
site $(y=0)$; see below.

\begin{figure}
{\centering
\resizebox*{3.3in}{!}{\includegraphics*{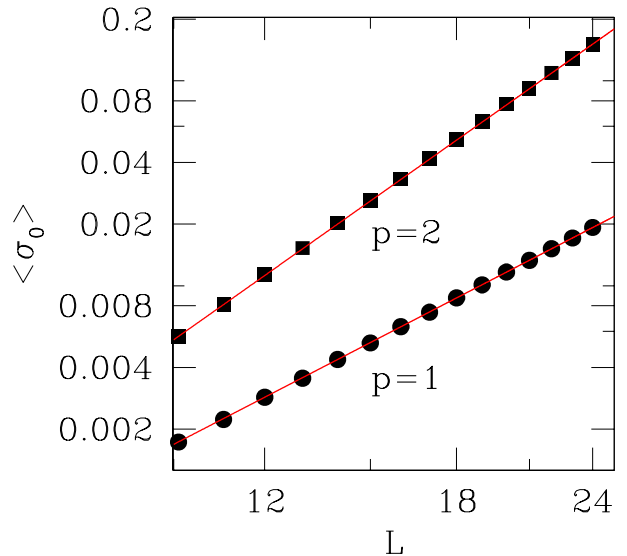}}}
\caption{(Color online)
Double-logarithmic plot of magnetizations at strip center for $p=1$ and $p=2$, against
strip width $L$. Here, $h_0=10^{-5}$. The straight lines are guides to the eye, with
respective slopes $2.75$ ($p=1$) and $3.75$ ($p=2$); see Eq.~(\protect{\ref{eq:int-a}}).
}
\label{fig:lfpower}
\end{figure}

The corresponding  predictions for $\langle \sigma_0\rangle$ have been checked against TM
results. In Fig.~\ref{fig:lfpower} we show $\langle \sigma_0 \rangle$ against strip width
$L$ ($10 \leq L \leq 24$), for $p=1$ and $2$, and $h_0=10^{-5}$ in
Eq.~(\ref{eq:field_def}), which amounts to having $\ell=10^5$ and $316.2$ respectively
for $p=1$ and $2$. Single power-law fits to TM data give $p+2-\eta=2.752(9)$ ($p=1$), and
$3.70(3)$ ($p=2$), illustrating that the agreement with Eq.~(\ref{eq:int-a}) indeed
improves as the ratio $L/\ell \to 0$.

\begin{figure}
{\centering \resizebox*{3.3in}{!}{\includegraphics*{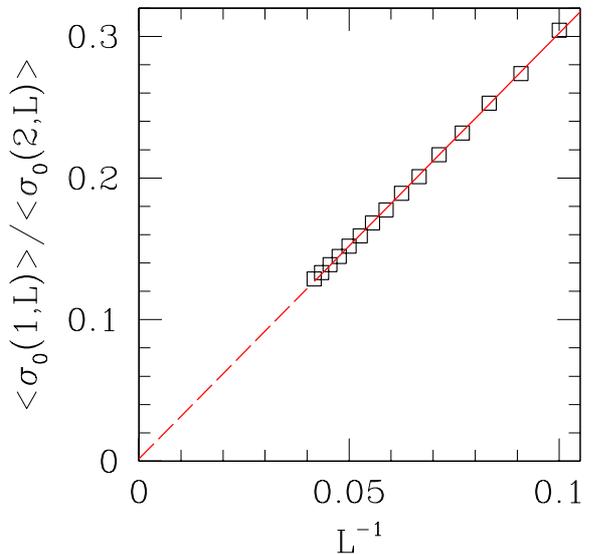}}}
\caption{(Color online) Plot of the ratio of magnetizations on the trapping axis,
$\langle \sigma_0(p_1,L) \rangle/\langle \sigma_0(p_2,L) \rangle$ against $L^{p_1-p_2}$
for $10 \leq L \leq 24$, $p_1=1$, $p_2=2$, with $h_0=10^{-5}$. Points are TM results; the
line is a linear least-squares fit of data (see text). }
\label{fig:lfratio}
\end{figure}

Furthermore, one can test the amplitude ratio ${\cal A}(p_1)/{\cal A}(p_2)$, by plotting
$\langle \sigma_0(p_1,L) \rangle/\langle \sigma_0(p_2,L) \rangle$ against $L^{p_1-p_2}$.
Our results for $p_1=1$, $p_2=2$ are shown in Fig.~\ref{fig:lfratio}. TM data are
well-adjusted by a linear least-squares fit ($y=ax+b$) with $b=0.001(1)$, providing
complementary evidence in favor of the single power-law scenario of Eq.~(\ref{eq:int-a}).
The adjusted slope is $a=3.01(2)$, to be compared to the predictions of
Eqs.~(\ref{eq:a1}) and~(\ref{eq:a2}),  ${\cal A}(1)/{\cal A}(2)= 2.98(1)$, where the
uncertainty follows from allowing $\alpha$ of Eq.~(\ref{eq:a2}) to vary between $1.0$ and
$1.5$.

We conclude that the linear response approach, in conjunction with conformal invariance
concepts, is a suitable description of the low-field magnetization properties of the
trapping problem on a strip geometry.

Our next step is to investigate the magnetization profiles across the system. If the
origin of coordinates in the integrals is now set at a distance $y_1$ from the trapping
axis, $0 < y_1 \leq L/2$, and defining the reduced coordinate $u \equiv y_1/L$, one can
see that the long-distance contribution $A_1(u,p)$ to ${\cal A}(u,p)$ [$\equiv {\cal
A}(u)$, for short] is still given by Eq.~(\ref{eq:a1}).
The short-distance part $A_2(u)$ can no longer be analytically expressed
in  a relatively simple form,  because the symmetries used in establishing 
Eq.~(\ref{eq:a2}) no longer hold off-axis.
However, one can still evaluate the corresponding integrals
numerically, in which case the full form of Eq.~(\ref{eq:conf-inv}) can be used over the
whole domain of integration. We have found that, in general:

(i) the order of magnitude of $A_2(u)$ remains similar to that of $A_1$, as is the case
on the strip axis [see Eqs.~(\ref{eq:a1}) and ~(\ref{eq:a2})];

(ii) the fractional variation $({\cal A}(1/2)-{\cal A}(0))/{\cal A}(0))$ from axis to
edge increases with $p$, from $\approx 2.5\%$ for $p=1$ to $\approx 6\%$ for $p=8$;

(iii) as suggested by Eq.~(\ref{eq:int-a}), together with generic scaling ideas, one gets
good scaling plots of TM results, $L^{-(p+2-\eta)}\,\langle \sigma_{y_1} \rangle$,
against $u$.

\begin{figure}
{\centering\resizebox*{3.3in}{!}{\includegraphics*{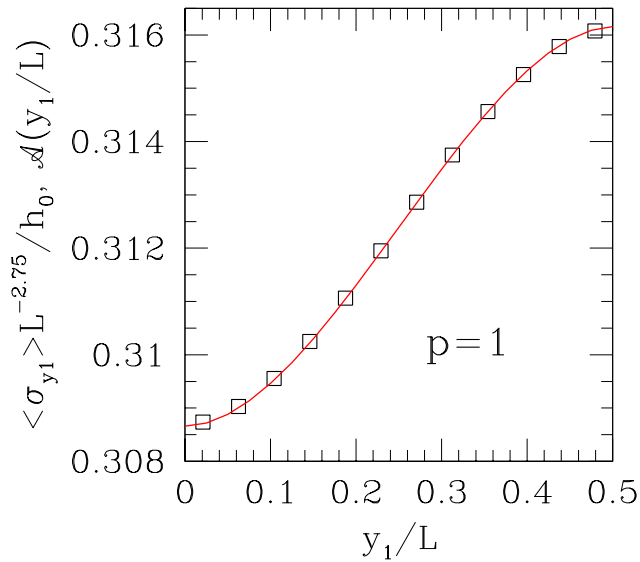}}}
\caption{(Color online) Scaled
magnetization TM data, $h_0^{-1}\,L^{-(p+2-\eta)} \langle \sigma_{y_1}\rangle$, for
$L=24$ (points), together with the result of linear response plus conformal-invariance
approach (solid line). Both for $p=1$. The horizontal axis is $u=y_1/L$. }
\label{fig:cprof_p1}
\end{figure}

We first illustrate the suitability of the linear-response plus conformal-invariance
approach, in the description of numerical TM data for magnetization profiles.
Figure~\ref{fig:cprof_p1} shows scaled magnetization data for $p=1$, $L=24$, as well as
${\cal A}(u)$ calculated by numerical integration.
Apart from an overall normalization factor for ${\cal
A}$, which affects only the vertical scale, there are no fitting parameters. One sees
that the agreement is very good.

\begin{figure}
{\centering \resizebox*{3.3in}{!}{\includegraphics*{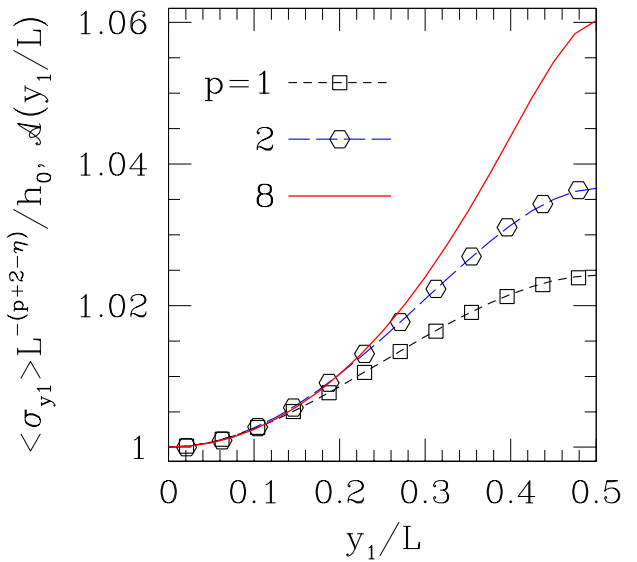}}}
\caption{(Color online) Calculated profiles ${\cal A}(u)$ for $p=1$, $2$, and $8$. Points
are scaled magnetization TM data, $h_0^{-1}\,L^{-(p+2-\eta)} \langle
\sigma_{y_1}\rangle$, for $p=1$ and $2$, both for $L=24$. The horizontal axis is
$u=y_1/L$. All curves have been normalized to unity at $u=0$. }
\label{fig:comp128}
\end{figure}

As regards point (ii), Fig.~\ref{fig:comp128} provides a visual comparison of calculated
profiles ${\cal A}(u)$ for $p=1$, $2$, and $8$, together with TM results for $p=1$ and
$2$ (the latter, both for $L=24$). In order to exhibit fractional variations more
clearly, all curves have been normalized to unity at $u=0$. Note that, for $p=2$, the
agreement between TM data and the respective calculated profile is as good as that for
$p=1$. TM calculations for the low-field regime $L/\ell \ll 1$ with $p=8$ have proved
rather unwieldy, as the corresponding fields on strip sites become numerically extremely
small, thus we do not present a comparison between the  calculated profile and TM data
for this case. In this connection, one must recall that the physically more interesting
trap shapes correspond to $p=1$ and $2$, since $p$ large approaches an infinite well.

\begin{figure}
{\centering\resizebox*{3.3in}{!}{\includegraphics*{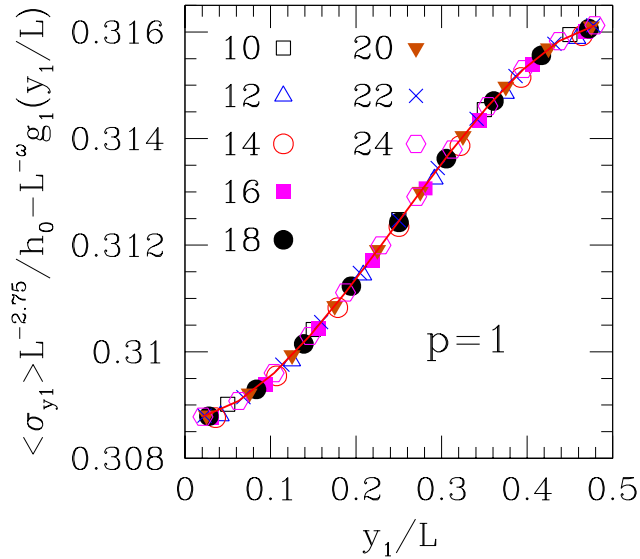}}}
\caption{(Color online) $p=1$:
data collapse for scaled magnetization TM results, including corrections to scaling,
i.e., $h_0^{-1}\,L^{-2.75}\langle \sigma_{y_1}\rangle -L^{-\omega}g_1(y_1/L)$ [see
Eq.~(\protect{\ref{eq:scale-corr}})], for $10 \leq L \leq 24$ (even widths only). The
horizontal axis is $u=y_1/L$. }
\label{fig:mprof_coll}
\end{figure}

For the data collapse mentioned in point (iii), we found that strong even-odd
oscillations are present, thus in the following we restrict ourselves to even lattice
widths. We also consider only $p=1$, which gives the smoothest results, as expected from
the fact that this setup corresponds to a shallow trap shape. Finally, corrections to
scaling are present which are small but noticeable for small and intermediate strip
widths. These have been accounted for in the usual way~\cite{dq09}, by assuming
\begin{equation}
h_0^{-1}\,L^{-(p+2-\eta)}\,\langle \sigma_{y_1} \rangle=f_p(y_1/L)+
L^{-\omega}\,g_p(y_1/L)\ .
\label{eq:scale-corr}
\end{equation}
Here, $\omega>0$ is the
exponent associated to the leading irrelevant operator. We found that, with $\omega=1$,
and $g_1(y_1/L)=c$ (constant), the best data collapse is obtained for $c=-7.0(2) \times
10^{-5}$. This is shown in Fig.~\ref{fig:mprof_coll}.

\section{High-field regime} \label{sec:hf}

In order to make contact with previous work~\cite{cv09}, we have set $p=2$, and generated
data for strips with varying $L$ and $\ell$, initially keeping constant the ratio
$\ell/L=1/\sqrt{2}$.

The average energy per site $e_s$ can be calculated, for a system with nearest-neighbor
interactions on a square-lattice geometry, as $e_s=e_x+e_y$, where $e_x$, $e_y$ are the
energies of the bonds along each coordinate axis. On a strip of width $L$ sites with PBC
across, with the TM advancing along $x$, one finds for the Ising spin-$1/2$ model at
$T_c$ and in the absence of (trapping or uniform) field, that $e_{x,y}=e_b^0 +
a_{x,y}\,L^{-2} +b_{x,y}\,L^{-4} + {\cal O}(L^{-6})$.  Sequences of $e_x$, $e_y$
generated using $4 \leq L \leq 16$ give $a_y/a_x=b_y/b_x=-1$ to four significant digits
($a_y,b_y >0$). Both extrapolate to the exact value $e_b^0=\sqrt{2}/2$ to four
significant digits, while the average between the two extrapolates agrees with that to
six digits.

Upon inclusion of the trapping field as described above, its value at sites near the
strip edge becomes rather large. The site energies now vary with position across the
strip. Since signatures of a scaling regime should be more apparent where deviatons of
the relevant fields from their critical values are small, at least in a local sense, we
calculate energies for the sites at, or nearest to, the strip center. A similar reasoning
appears to have been followed in the Monte Carlo simulations of Ref.~\onlinecite{cv09}.

The resulting sequences of $e_x$ and $e_y$ exhibit even-odd oscillations. However, these
are smoothed out when $e_s$ is considered instead. Fitting data for $10 \leq L \leq 23$
to $e_s=\sqrt{2}+a\,\ell^{-\theta}\,(1+b\,\ell^{-2})$ gives $\theta=0.514(10)$, in very
good agreement with the scaling prediction~\cite{cv09}, $\theta=16/31=0.51613 \dots$ .

Next, we consider local magnetizations $m_0$, again on sites at, or nearest to, the strip
center, for the same reasons invoked above in regard to site energies. The magnetization
sequences show significant even-odd oscillations, thus we ran separate fits for data
subsets corresponding to even and odd $L$. Using $10 \leq L \leq 23$, assuming
$m_0=c\,\ell^{-\phi}\,(1+d\,\ell^{-2})$ we get $\phi=0.0646(8)$ from an average between
the final estimates for even- and odd- sequences. This is again in very good agreement
with the respective scaling prediction~\cite{cv09}, $\phi=2/31=0.0645 \dots$ .

\begin{figure}
{\centering \resizebox*{3.3in}{!}{\includegraphics*{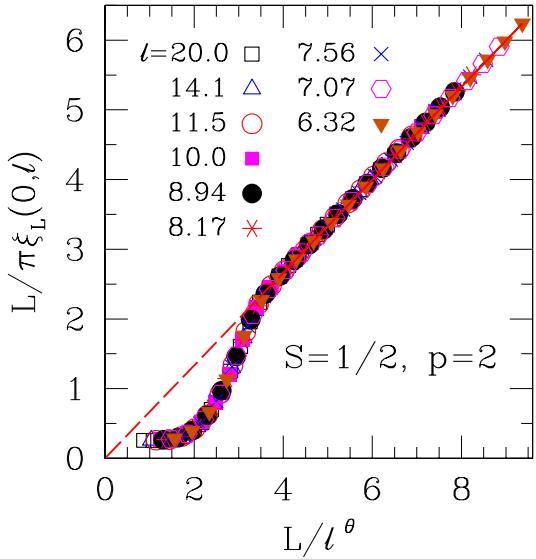}}}
\caption{(Color online) Scaled correlation-length data for $p=2$, with varying $L/\ell$
[see Eq.~(\protect{\ref{eq:xi_scale}})]. The horizontal axis is $x \equiv L/\ell^\theta$,
with $\theta=0.51$. The straight line through the origin is a guide to the eye. }
\label{fig:parcl}
\end{figure}

In analogy with Eq.~(\ref{eq:fss_fe2}), and considering the uniform field $H \equiv0$
from the start, the correlation length on a strip is given by:
\begin{equation}
\xi_L(t,\ell)=L\,f(Lt,L\,\ell^{-\theta})
\label{eq:xi_L}
\end{equation}
At $t=0$, and
applying standard FSS ideas, one expects:
\begin{equation}
\frac{L}{\pi\,\xi(0,\ell)}=g\left(\frac{L}{\ell^{\theta}}\right)\ ,\ \ g(x) \sim
\begin{cases}{{\rm const.}\quad x \ll 1}\cr{x \qquad\quad x \gg 1} \end{cases}\ .
\label{eq:xi_scale}
\end{equation}
We generated correlation-length data for $4 \leq L \leq 24$ and assorted $6 \lesssim \ell
\lesssim 20$, thus allowing $L/\ell$ to vary over a relatively broad extent. Recalling
from conformal-invariance~\cite{cardy} that $L/\pi\,\xi_L(0,0) =\eta=1/4$, we plotted
$L/\pi\,\xi_L(0,\ell)$ against $L\,\ell^{-\theta}$ for tentative values of $\theta$,
searching for the best data collapse. This was found along a somewhat broad range, $0.49
\lesssim \theta \lesssim 0.52$, which includes both the trap-size scaling prediction and
the mean-field value $\theta_{\rm MF}=1/2$. Fig.~\ref{fig:parcl} shows a typical result,
using $\theta=0.51$. The trends predicted in Eq.~(\ref{eq:xi_scale}) are all verified.

In order to narrow down our error estimates, we investigated how well the 
large-$x$ portion of the data falls on a straight line, as predicted in 
Eq.~(\ref{eq:xi_scale}). We fitted portions of data contained within intervals 
$x_1 < x < x_2$ to a single power law with an adjustable exponent, $\varepsilon$. 
For $[x_1,x_2]=[4,9]$, i.e. essentially all the large-$x$ region, we found 
$\varepsilon=1$ within $0.5\%$ for $\theta=0.503(7)$, while if we restricted 
ourselves to $[x_1,x_2]=[4,6]$ the corresponding range of $\theta$ turned out to 
be $\theta=0.514(10)$.

Since the largest values of $x$ correspond to large $\ell^{-1}$, the respective
correlation-length data belong to a region in parameter space away from the critical
point, where the scaling behavior crosses over to a mean-field picture. Accordingly, when
one includes such data in the analysis, the results for $\theta$ are biased towards its
mean-field value; on the other hand, their exclusion brings the estimate of $\theta$ back
towards the scaling prediction.

An analytic treatment of the correlation length can be given for the limit $p \to
\infty$, providing for this case the scaling function
$g(x)$ in Eq.~(\ref{eq:xi_scale}), and having similarities to the form shown for
$p=2$ in Figure~\ref{fig:parcl}. The reduction for $p \to \infty$ comes because the
trapping well is there equivalent to confinement between $-\ell$ and $+\ell$,
so strip geometry and well are together equivalent to a strip of width
$L^\prime={\rm min}(L,2\ell)$. For this equivalent system,
Eq.~(\ref{eq:conf-inv}) (with $L$ replaced by $L^\prime$) can be used to obtain the
effective correlation length, yielding for $p \to \infty$,
\begin{equation}
\frac{L}{\pi\,\xi_L(0,\ell)}=g\left(\frac{L}{\ell^\theta}\right)\ ,\ {\rm with}\
g(x)=\frac{\eta}{2}\,{\rm max}(2,x)\
\label{eq:p_infty}
\end{equation}
[using $\theta (p=\infty)=1$, from Eq.~(\ref{eq:theta})].

\section{Universality: Ising $S=1$ and Blume-Capel model} 
\label{sec:bc}

In this section, we investigate the universality of trapping scaling
against varying features of the spin Hamiltonian. We shall keep
$p=2$ in what follows. For TM calculations, the field setup is the
same as defined in Sec.~\ref{sec:basics}, with the trapping axis
halfway along the strip width.

We start by applying the ideas discussed in the previous Section to an
Ising $S=1$ system, on a square lattice with nearest-neighbor couplings,
in units of which the zero-field critical temperature is $T_c=1.69356(2)$~\cite{bcg03}.
With the trapping field coupled to the magnetization, we considered the
high-field regime, making again $\ell=L/\sqrt{2}$ as our starting point.

At the zero-field critical point, by considering the
bond energies $e_x$ and $e_y$ as in Sec.~\ref{sec:hf}, we found from fits of data
for $4 \leq L \leq 13$ that $e_s^0=1.16100(2)$, to be compared with
$e_s^0=1.16094(5)$~\cite{bn85}. Upon introducing the trapping field,
we restricted ourselves to the sites closest to the strip center,
for the reasons invoked in Sec.~\ref{sec:hf}. Fitting data for $4 \leq L \leq 15$
(even- and odd-$L$ sequences have to be considered separately, on account
of the associated oscillations) to $e_s=e_s^0+a_1\ell^{-\theta}(1+b_1\ell^{-2})$
gives $\theta=0.55(2)$, where the error bar reflects the spread among fits of
differing subsets of data: odd/even, and 
vertical/horizontal/vertical-plus-horizontal bond energies.

For the local magnetization closest to the strip center, for $4 \leq L \leq 15$
we again resorted to separate fits for even- and odd-$L$ sequences. The relative
shortness of such sequences, compared to the $S=1/2$ case, makes extrapolation
somewhat risky. However, by attempting power-law fits $m_0=c_1\ell^{-\phi}(1+d_1
\ell^{-2})$, we saw that, by successively excluding an increasing number of
smaller strips widths, the sequences of estimates for $\phi$ vary as:
$\phi=0.088(4)$, $0.075(2)$, $0.070$ ($L$ even); and $\phi=0.099(6)$,
$0.078(2)$, $0.071(2)$, $0.068$ ($L$ odd). Thus, it seems plausible
that consideration of larger $L$ would produce a result compatible
with the prediction $\phi=2/31$~\cite{cv09}.

Next, by allowing the trapping field intensity (thus, the corresponding
length $\ell$) to vary independently of $L$, we extracted an independent
estimate of $\theta$ via the scaling argument presented in Eqs.~(\ref{eq:xi_L})
and~(\ref{eq:xi_scale}). We used $4\leq L \leq 15$, and
$2 \lesssim \ell \lesssim 9$. For tentative values of $\theta$ close to $0.5$,
we found a scaling curve quantitatively very similar to that for $S=1/2$,
shown in Fig.~\ref{fig:parcl}.  Indeed, the respective numerical values
stay within at most $3-4\%$ from those of the $S=1/2$ curve.
However, the quality of data collapse is markedly inferior; furthermore,
it does not appear possible to have good  scaling with a single $\theta$
for the whole curve.
For $\theta=0.47$, points fall very well on a straight line at large $x=L/\ell^\theta$,
while those at smaller $x$ show a somewhat large degree of scatter. For $\theta=1/2$
the situation is reversed.

Overall, the above results are in fair agreement with the hypothesis of universal
behavior against spin value, though with some degree of scatter. 
Note that the estimates for $\theta$ from
internal-energy and correlation-length scaling are, respectively, above and below
the expected value $\theta=16/31$; thus, we have found no systematic trend
away from universality. 

We now turn to discussing the application of a trapping field to a variant
of the $S=1$ Ising model, namely the Blume-Capel (BC)
model~\cite{b66,c66,beale,xapp98,dgb05}, on a two-dimensional square lattice. The
Hamiltonian is:
\begin{equation}
{\cal H} =-J \sum_{(i,j)} S_i S_j +\left(D+h_0\,|{\vec
r}_i|^p \right) \sum_i S_i^2\ ,
\label{eq:bcdef}
\end{equation}
where $S_i=\pm 1,0$
(spin--$1$ Ising variables). With $J>0$ (taken to be unity from now on) and $D, h_0>0$,
the field term privileges vacancies, so there is an immediate mapping between $\langle
S^2 \rangle$ and the "particle" density in the lattice-gas picture. The BC model was
studied recently via Monte Carlo simulations, in a trapping field with $p=1$, and
assorted values of $D$~\cite{pbs08}.

Considering for now $h_0=0$, it is known that for $D>0$ the competition between field and
spin-spin coupling leads to a tricritical point at $T_t=0.6085776(1)$,
$D_t=1.9658149(2)$~\cite{beale,xapp98,dgb05}. For $D<D_t$ the behavior along the $D-T$
critical line is governed by the Ising spin--$1/2$ fixed point at $D \to -\infty$ (at
which vacancies are effectively forbidden)~\cite{xapp98,dgb05}.

Therefore, as regards the respective universality class of the (suppressed) phase
transition, by fixing $D<D_t$ and taking $h_0 \neq 0$ one is in fact crossing the
critical line far away from its governing fixed point. Though this does not invalidate
the conclusions of Ref.~\onlinecite{pbs08}, which concern the validity of the
local-density approximation for calculating the internal energy and linear structure
factor, one must expect strong crossover effects if attempting to evaluate critical
indices there.

In what follows, we kept $D=0$, and $p=2$ in Eq.~(\ref{eq:bcdef}).

 The "order parameter" associated to the trapping
transition, i.e. the vacancy concentration, can be extrapolated from TM data for $4 \leq
L \leq 13$ to be $1-\langle S^2\rangle =0.16120(1)$, in excellent agreement with the
estimate from Table IV of Ref.~\onlinecite{dgb05}, namely $\rho({\rm Ising}) =
0.1610(5)$. This is a reminder that one is working away from the critical point. Thus,
for instance, in this case it is not possible to define the analogue of the magnetization
exponent $\phi$ of Sec~\ref{sec:hf}.

\begin{figure}
{\centering \resizebox*{3.3in}{!}{\includegraphics*{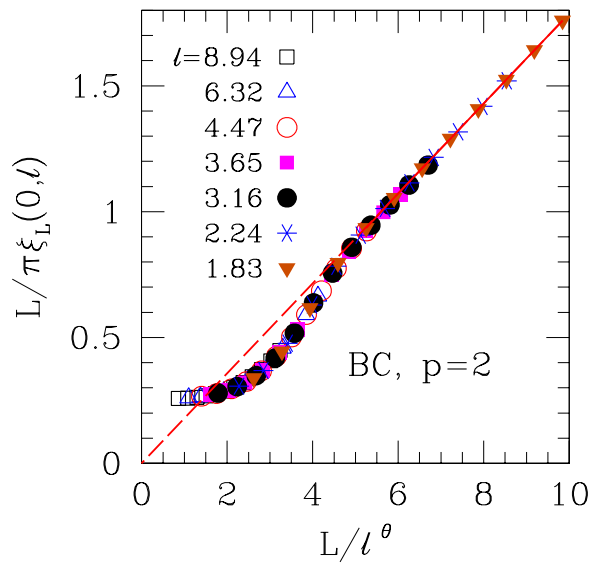}}}
\caption{(Color online) Scaled correlation-length data for BC model, $p=2$, with varying
$L/\ell$ [see Eq.~(\protect{\ref{eq:xi_scale}})]. The horizontal axis is $x \equiv
L/\ell^{\theta_{\rm app}}$, with $\theta_{\rm app}=0.700$. The straight line through the
origin is a guide to the eye.}
\label{fig:parbccl}
\end{figure}

On the other hand, since for sufficiently small $h_0$ the associated length $\ell$ still
diverges, at $t=0$ one can examine the interplay between $\ell$ and, say,
$\xi_L(0,\ell)$, and find the respective scaling behavior and (apparent) exponents. With
this proviso, the arguments leading to Eq.~(\ref{eq:xi_scale}) are still valid. Using $4
\leq L \leq 15$, and $2 \lesssim \ell \lesssim 9$, we attempted data collapse of
$L/\pi\xi_L(0,\ell)$ against $L/\ell^{\theta_{\rm app}}$, with the best superposition for
$\theta_{\rm app}=0.700(5)$. The results are shown in Fig.~\ref{fig:parbccl}.
Comparison of the respective numerical values shows that this scaling curve
exhibits no superposition (except for the limiting constant value of $1/4$
at $x \to 0$) with that for the $S=1/2$ Ising case, shown in Fig.~\ref{fig:parcl}.

In a similar vein, we investigated the site energies.
Making now $\ell/L=1/\sqrt{2}$, adjusting a sequence of
data for $4 \leq L \leq 15$ to $e_s=e_s^0+a^\prime \ell^{\,\theta_{\rm 
app}^\prime}$, we
get $0.10 \lesssim \theta_{\rm app}^\prime \lesssim 0.13$. The discrepancy between
$\theta_{\rm app}^\prime$ and $\theta_{\rm app}$ of the correlation-length scaling again
reflects the fact that one is not close to an actual fixed point.

\section{Discussion and Conclusions}
\label{sec:conc}

Thermodynamic systems with degrees of freedom subject to a trapping field bring 
yet another length scale 
into play, namely, that of the trap size, $\ell$. The latter is a measure of how 
far from the center the trapping field reaches a given intensity, $h_0=\ell^{-p}$, 
with $p$ describing the steepness; $p=2$ for the usual harmonic traps. Therefore, 
analyses of data extracted from systems of finite extent, $L$, (at least in one of its 
$d$ dimensions), must incorporate the interplay between these two independent 
length scales. We have proposed an extension of the usual finite-size scaling 
theory to deal with these problems, and tested it on classical (Ising) systems, 
for which the scaling limit of large $L$ can be easily achieved. The tests have been 
carried out on long strips of finite width $L$, subject to a 
trapping field translationally invariant along the "infinite" direction; the strip 
topology was that of a square lattice.

For all field regimes considered, the characteristic exponents of trap scaling
vary continuously with $p$. This is analogous to what happens in the 
two-dimensional $XY$ model, which has a line of fixed points, 
and exponents varying along the line~\cite{nelson}. 
There, the position on the line is related to a relevant variable
which is marginal in the sense that it does not remove criticality,
although it does affect exponents. Here we have the same scenario (but
without, yet, a flow diagram to support it).

In the low-field regime (i.e., for "shallow" traps, $\ell\gg L$) one expects 
a linear-response scenario to be appropriate. For Ising systems we 
can therefore use this, together with exact conformal-invariance results,
to produce analytical predictions to be compared with numerical results from TM 
methods, at the critical point of the  otherwise infinite and untrapped system. 
We have established that, in accordance 
with our {\it ansatz}, the local magnetization displays the scaling behavior
\begin{equation}
\langle \sigma_{y_1} \rangle =h_0 \,L^{(p+2-\eta)}\, f_p(y_1/L),
\label{eq:scale-leading}
\end{equation}
to leading order in $h_0$, where $y_1$ is the distance from the center along the 
finite strip direction; see Sec.~\ref{sec:lowfield}.

In the high-field regime, $\ell\lesssim L$, scaling bears the signature of the 
$p$-dependent trap exponent $\theta$, as previously discussed in Ref.~\onlinecite{cv09}
[see Eq.~(\ref{eq:theta})]. For $p=2$ we have provided independent checks of the exponents
for the  scaling at criticality of the central magnetization and of the energy density,
in very good numerical agreement with Ref.~\onlinecite{cv09}. Furthermore, with our 
TM calculations over a broad range of values of $\ell$ and $L$ 
(spin-1/2), we have established  that the scaled correlation lengths behave as 
$L/\pi\,\xi_L(\ell)\sim g(L/\ell^\theta)$, where the function $g(x)$ very accurately
follows a form predicted by scaling arguments. Controlling for the quality of 
data collapse provides us with an estimate for the exponent $\theta \approx 0.51$
(for $p=2$) which agrees very well with the respective scaling prediction, 
$\theta=16/31$. We have also given an analytic treatment which describes
how the finite-$p$ scaling function evolves, in the $p \to \infty$ limit,
towards a piecewise straight-line shape [see Eq.~(\ref{eq:p_infty})].

As a further check of the theory, we have also considered the high-field regime 
for a spin-1 Ising model in two situations. The first corresponds to an immediate 
extension of the spin-1/2 case, in which the trapping field couples to the 
magnetization. Since the  TMs for $S=1$ are larger than for $S=1/2$, 
we were not able to consider linear lattice sizes as large as previously.
Thus, the quality of the data collapse and scaling fits
was somewhat compromised; however, 
our results are broadly consistent with fits to the same exponents as before, namely, 
$\theta=16/31$ and $\phi=\theta\beta/\nu=2/31$ (in standard notation of critical 
exponents) for $p=2$. This picture therefore gives support to the universality of 
trapping exponents against spin magnitude. 

We have also  examined the Blume-Capel (BC) model, with the trapping field 
coupling to $\sum_i S_i^2$ 
(whose average gives the "particle density", in the lattice-gas 
language). This latter feature might be considered a more transparent 
way to connect the spin language to that of particle trapping.
However, we have seen that the fixed-point structure of the BC phase diagram 
puts one at a disadvantage, concerning
the estimation of scaling exponents for physically plausible values of the
model's parameters. Notwithstanding this, we have been able to show that
the competition between the associated scaling fields, $L^{-1}$ and $\ell^{-1}$,
gives rise to an effective scaling picture in qualitative agreement with
standard FSS ideas.

\begin{acknowledgments}
S.L.A.d.Q. thanks the Rudolf Peierls Centre for Theoretical
Physics, Oxford, where parts of this work were carried out, for hospitality, 
and CNPq for funding his visit. 
R.R.d.S. is grateful to R.T. Scalettar for interesting discussions on this  problem.
S.L.A.d.Q. and R.R.d.S. acknowledge joint financial support by grants from the 
Brazilian agencies FAPERJ (Grants No. E26--100.604/2007 and No.
E26--110.300/2007) and CAPES; 
S.L.A.d.Q. and R.R.d.S also hold individual grants from the Brazilian agency 
CNPq (Numbers 30.6302/2006-3 and 31.1306/2006-3, respectively).
R.B.S. acknowledges partial support from EPSRC Oxford
Condensed Matter Theory Programme Grant EP/D050952/1.
\end{acknowledgments}

\end{document}